# PERMEATION AS A DIFFUSION PROCESS

November 18, 2005

Bob Eisenberg
Dept. of Molecular Biophysics and Physiology
Rush Medical Center
1750 West Harrison Street
Chicago Illinois 60612
USA

***Introduction.*** The description of ionic movement through biological membranes is an old topic, nearly as old as the description of diffusion itself. For some 150 years, the movement of molecules through membranes has been known to depend on the nature of the membrane and the concentration of the molecules. For nearly as long, the movement of ions has been known to depend on the electrical potential across the membrane. The biologist has always been interested in the movement of molecules and ions and how that is modified by concentration, electrical potential, and controlled by the biological system, i.e., the tissues, cell, membrane, or channel. This interest has produced measurements of current and flux, under a range of concentrations and membrane potentials, and theories of various resolution to account for these, as described later in a historical section. Here we share the perspective (if not prejudice) of present day workers, emphasizing the resolution of modern biology, which is as atomic as it is molecular.

Ions are atoms and the ultimate description of their movement through biological membranes should be on the atomic level, with spatial resolution of atomic locations (i.e., 0.1 Å) and temporal resolution of atomic motions ($10^{-16}$ s). But most biological functions of membranes are macroscopic involving the electrical current on the time scale of micro- to milliseconds that is produced by the movement of ions through channels, each of which takes some $10^{-7}$ s to pass through from one side to the other. Any description of a channel relevant to biological phenomena must be on the biological time scale and contain the variables important to the function of cells; these are the variables manipulated and measured experimentally. The description must yield the current or flux through the membrane often measured in experiments and it must include the concentration of ions, the *trans*membrane potential, and the structure of the channel and membrane at the highest resolution possible.

Atomic simulations are the obvious and most appealing way to describe ion movement through channels and membrane but it is easy to overlook the difficulties of simulating the macroscopic variables (e.g., concentration, potential, and current) known to be important experimentally and biologically. The next pages are meant to demonstrate the inherent difficulties of simulations at atomic resolution, and go on at some length, because of the popularity and evident appeal of such calculations. The reader who is ready for a macroscopic treatment can jump ahead to the section labeled *Averaging*.

**<u>*Simulation of Concentration*</u>**. The difficulty in achieving atomic resolution and biological relevance in a single description of ionic movement is illustrated clearly when we try to simulate a concentration. It is obvious that we must simulate concentration if we are to describe the movement of ions through channels because diffusion and permeation depend on concentration.

Consider a typical Ringer solution surrounding a mammalian cell. An atomic detail computation must estimate the concentration of all the ions present, typically including $Na^+$ (~150 mM, often the most concentrated cation present) and $Ca^{++}$ (~2 mM, often the least concentrated, not counting $H_3O^+$). Enough ions must be present of each so that statistical uncertainties in the concentration are reasonable. To keep numbers round, let us accept an uncertainty of 1%. If the ions behave independently, some 10,000 ions must be in the calculation, because errors (of independent identically distributed random variables) go as (the reciprocal of) the square root of the number of variables. In fact, ionic motion is highly correlated because the electric field generated by the charge of one ion influences the motion of other ions for substantial distances and times, but we shall ignore that correlation for the purposes of this calculation. Considering correlations would increase, perhaps dramatically, the estimate of the number of particles needed in simulations.

Ions in solution are accompanied by water, in fact, 1 liter of solution contains ~55 moles of water, each mole of water containing 1 mole of oxygen and two moles of hydrogen. Thus, for each $Na^+$ ion in a 0.15 molar solution, there are 55/0.15 water molecules. To calculate the properties of 10,000 $Na^+$ atoms, one must calculate the properties of the $3 \times (55/0.15) \times 10^4 = 1.1 \times 10^7$ atoms surrounding them. Similarly, to calculate the properties of 10,000 $Ca^{++}$ ions in a 0.002 molar solution, one must also calculate the properties of $3 \times (55/0.002) \times 10^4 = 8.3 \times 10^8$ atoms. It is very difficult to perform simulations of this size and is likely to remain so for many more iterations of Moore's law. (Moore's law is the empirical observation that the speed and size of computations have doubled approximately every 1.5 years for many decades.) We include all the atoms of the water because the internal motions of water are calculated in present day simulations of molecular dynamics of proteins and solutions. There is no generally agreed upon way to avoid calculating the internal dynamics of the atoms of water and thereby replace the expensive calculation of the dynamics of water with the properties of an averaged molecule.

The situation is more extreme when considering intracellular solutions. In these, ions are often present that have dramatic effects on channels at very low concentrations. These ions often cannot be left out of the solution without drastically changing channel properties; indeed, in some cases, if the ions are left out, the channel no longer functions at all. These ions might be called co-factors in analogy to the coenzymes so important for many other proteins (Eisenberg, 1990).

Most notably for our purposes, intracellular $Ca^{++}$ has a crucial role controlling many biological systems at concentrations $< 10^{-6}$ M. Many channels that have at least one end in the intracellular solution are very sensitive to tiny intracellular concentrations of calcium and do not function normally if that calcium is omitted. Simulation of a micromolar $Ca^{++}$ solution requires calculation of $3\times(55/10^{-6}) \times 10^4 = 1.65\times10^{12}$ atoms. Calculations of this size are difficult and are likely to remain so, no matter what the size of computers, given the realities of the finite word length and round off error inherent in any computer calculation.

**_Simulations of current_**. The difficulty in simulating concentration arises fundamentally because of the gap between atomic dimensions needed to resolve molecules and the macroscopic dimensions needed to define concentration. The difficulty in the calculation of current arises because of the gap between atomic time scales (needed to see molecular motion) and the macroscopic time scale (needed to see biological function).

The biological functions of channels occur on time scales longer than 10 microseconds, and even the time for permeation of a single ion is some 100 nsec (one picoamp corresponds to one elementary charge every 160 nsec in a channel always occupied by one ion). Simulations of molecular dynamics are nearly always done with time steps of $10^{-16}$ sec so that atomic vibrations can be resolved. Thus, calculations take some $10^9$ time steps before a single ion crosses a channel, assuming they always move in one direction. To estimate a current with 1% reliability, one needs some $10^4$ crossings of ions, if the crossings are independent of each other in the stochastic sense, meaning that at least $10^{13}$ calculations are needed to estimate the current flowing through a channel at a single *trans*membrane potential and in a particular pair of solutions. Brownian motion guarantees an enormous number of changes in direction, ignored in this estimate, and ions in multiply occupied channels are unlikely to move independently. Thus, the number of calculations actually needed is very much larger than $10^{13}$. Reliable calculations of this many time steps are difficult and are likely to remain so, for some time, given the realities of the finite word length and round off error inherent in any computer calculation.

**_Simulations of potential_**. Problems involved in calculating the electric field are similarly difficult. The electrical potential $\phi(r; r_k)$ (units: V = J/C) at location $r$ produced by a set of charges $q_k$ (units: C) at locations $r_k$ is determined by Coulomb's law, the fundamental relation between charge and potential that in essence sums the effect of each atomic charge.

$$\phi(r; r_k) = \frac{1}{4\pi\varepsilon_r\varepsilon_0} \sum_k \frac{q_k}{r - r_k} \tag{1}$$

$\varepsilon_r$ is the (relative) dielectric constant (no units, value approximately 80 for distilled water), $\varepsilon_0$ the permittivity of free space ($8.85 \times 10^{-12}$ F/m). Symbols and units are particularly diverse in physical chemistry and so I try to follow the suggestions of the official bodies of physical chemistry (Physical Chemistry Division of the International Union of Pure and Applied Chemistry: Mills, 1988) although this sometimes means using unfamiliar units (e.g., $dm^3$ for liter) or symbols.

Coulomb's law is as exact as any physical law (Feynman, Leighton and Sands, 1963) but using it requires a knowledge of all charges in the system, because in many cases even small amounts of charge create large effects.

The distribution of charge is often hard to determine, because it depends on many variables and often in complex time dependent ways. For example, in most materials the movement of a charge changes the distribution of other charge, just as the presence of a star changes the distribution of matter in a nearby star, and that change is often significant.

The charge induced by a nearby charge is customarily described by a dielectric constant, but it is important to understand the depth of the approximations involved. Most matter, and all ionic solutions, contain charge that moves significantly when a test charge is placed nearby. The charge movement is usually significant, because the forces that hold matter are fundamentally electrical of more or less the same strength as the electric field produced by the test charge. Understanding how charge moves in a material in response to an applied electric field, or a test charge introduced into the material, is a central topic in material science. Those charge rearrangements are often responsible for technologically important properties of matter. In ionic solutions, there are many types of charge movement in response to an electric field both in the solvent and solute. The solvent water molecule rotates in the electric field, and can even be displaced, although the water molecule is net neutral, if the electric field has a nonzero second (spatial) derivative. The ions of the solution move as well, and their movement is in fact responsible for many of the macroscopic properties of ionic solutions.

In simple cases, when electric fields are not large, time scales are longer than microseconds, and mobile ions are not present, the induced charge is proportional to the local electric field. In that case, equation (1) describes the potential, provided the charges on the boundary of the system are included. If ions are present, their movement in response to an electric field must be described as well, and later we shall see how we do that (eq. (6) and following pages).

**_Boundary conditions_**. Boundaries are very important in electric field problems because the potential on the boundary is often controlled by an active device that injects charge (e.g., current) into the system, as the system changes, for example, as molecules move in it. One example is the voltage clamp arrangement used widely for studying ion channels for more than 50 years. If the boundaries are ignored in such a system, the injected charges are also ignored and the qualitative properties of the system change (e.g., it is no longer voltage clamped). Sometimes these boundary charges are safe to ignore, because in some circumstances the electric field is a short range force; but in general this is not the case. Rather, the electric field can extend arbitrarily far, depending on the geometry and properties of its boundaries, as consideration of a 19$^{th}$ century telegraph should make clear.

When boundaries are important to a system, they are often connected to devices that allow the flow of current and charge. For example, systems are often "grounded" by a wire so that charge and current can flow into a grounded location in a system and keep it at a constant zero potential. No external energy source drives this flow; rather it is driven by the electric field created by the separation of charge between system and ground. In this case, it is difficult to keep track explicitly of the charge and current at the ground point and in the grounding wire.

It is often easier then to write a form of Coulomb's law that does not depend explicitly on the charge at the boundaries. Coulomb's law can be rewritten as a differential equation, Poisson's equation, in which the charge on the boundary need not be specified. Simply specifying the potential on the boundary (e.g., setting the potential equal to zero at a grounded location) is enough; the charge on the boundary is an output of the solution of the Poisson equation. It does not have to be known in advance. Textbooks like Feynman, *et al.,* 1963; Griffiths, 1981, and Jackson, 1975, show how to solve electric field problems with boundary conditions by solving Poisson's equation. Smythe, 1950, is a classic physics reference and Jack, Noble, and Tsien, 1975, is the classic physiological reference.

In one important case, boundaries are not directly involved in determining the electric field, namely the case when the electric field does not reach the boundary, and the boundary is uncharged. If we deal with an isolated charged molecule of radius *a*, in a uniform ionic salt solution, with *uncharged* boundaries very far from the charge of interest (so therefore with boundaries that are *not grounded*, nor voltage clamped, nor connected to anything else), the electric potential $\phi(R)$ spreads more or less exponentially on a characteristic scale of the Debye length $\kappa^{-1}$ according to the traditional Poisson-Boltzmann-Debye-Hückel treatment of ionic solutions (see p. 772-777 of Berry, Rice, and Ross, 2000)

$$\phi(R) = \frac{Z_i e}{4\pi\varepsilon_0 \varepsilon_r} \frac{e^{\kappa a}}{1+\kappa a} \frac{e^{-\kappa R}}{R} \qquad (2)$$

where

$$\kappa^2 = \frac{e^2}{\varepsilon_0 \varepsilon_r k_B T} \sum_k Z_k^2 \frac{N_k}{V} \qquad (3)$$

*R* is the distance to the center of the charged molecule (units: m); *e* is the charge on a proton (units: C); $\varepsilon_0 \varepsilon_r$ is the permittivity of the solution (units: $F/m$); $k_B T$ is the thermal energy (units: J); with $k_B$ the Boltzmann constant (units: $J/\deg$); and *T* the absolute temperature (units: deg K); $N_k/V$ is the density of a single species of ion *k* (e.g., $Na^+$); $Z_k$ is the number of charges per ion; and $N_k/V$ is the (number) density (e.g., "concentration") of ions; where $N_k$ are the number of ions in volume *V*.

It is useful to remember the approximate equation for the Debye length in a uni-univalent salt (like $Na^+Cl^-$) of concentration *C*

$$\kappa^{-1} \doteq \frac{3}{\sqrt{C}}; \quad \kappa^{-1} \text{ in Å, } C \text{ in M} = \text{mol/L} = \text{mol/dm}^3 \qquad (4)$$

where *C* is the concentration of monovalent ions surrounding the protein in molar units and $\kappa^{-1}$ is measured in Angstroms. Note the standard but relatively unknown unit "deci-cubic meter" $dm^3$ meaning one tenth of a cubic meter, i.e., 1 L (p. 38 footnote 12 of Mills, 1988).

If the concentration is 0.15 molar, as in a typical biological solution, $\kappa^{-1} = 8 \text{Å}$ and the potential drops below 1% of its peak value 15 Å away from the center of a $Na^+$ of radius 1.54 Å, according to eq. (2). Further than this distance, the $Na^+$ has no significant effect on potential in an isolated system. Further than this, the effect of the charge is said to be screened, or shielded from the rest of the system. Physically, what happens is clear. If a charge moves in an isolated system, neighboring charges move so they nearly compensate for the effects of the introduced charge in the sense that the spatial distribution of potential (> 15Å distance away) is hardly changed. Of course, *the insertion* of a new charge into an isolated system is different. It has an effect on the average potential of the entire system with respect to another system (e.g., with respect to ground). The system is no longer electrically neutral after a charge is added and that net charge changes the average potential, sometimes significantly (particularly when the system is small), even if its local effect is shielded.

If a charge is introduced into a system that has a boundary with defined properties, the situation is different, because of the charge on the boundary. For example, if the system were a beaker of salt solution containing an electrode connected to ground, or connected to a voltage clamp, there would be a long distance effect of the electrical field. If a charge is introduced into the salt solution, a ground wire, or a voltage clamp would maintain the potential at the electrode (i.e., boundary) by supplying an equal and opposite charge, *no matter how far the electrode is from the charge.* This example neatly illustrates how the electric field can be both short and long range, depending on the details of the situation.

If we neglect the long range components of the electric field, it seems easy to include enough ions to simulate an electric field. A sphere of radius 15Å of 0.15 molar salt solution contains about 1 ion, an eminently feasible number for simulations, if the system is isolated and no other charges are involved. At least it appears so until one considers time.

The time course of atomic motion introduces both deterministic and random effects. If a charge is moved in a solution, the surrounding charges take time to respond and this time is long compared to the time scales of atomic motion simulated in molecular dynamics. Each component of charge responds on its own time scale. Only electron movement in atomic orbitals is 'instantaneous' on the time scale of femtoseconds. For example, the shielding phenomena described by eq. (2) take very long indeed to develop compared to femtoseconds; typically, shielding phenomena of the Debye–Hückel type take nanoseconds to develop. Before then, the ions are unshielded (by their ionic atmosphere) and the qualitative properties of the system are different. Simulations which do not extend long enough to allow the ionic atmosphere to reach steady state are likely to reveal phenomena not directly relevant to biological permeation: biological permeation occurs on a time scale slower (not faster) than the relaxation of the ionic atmosphere. In the world of ion permeation, shielding phenomena are always present and probably always close to steady-state. Time dependence also introduces stochastic properties of some importance.

***Atomic motion***. Atoms are in continual motion, at thermal velocity, which is more or less the speed of sound in water, 1500 meters/sec, i.e., 15 Å every pico second. This motion ensures fluctuations in the number of ions in a region, and fluctuation in the number of charges produces fluctuation in potential.

Let us consider what would happen if one ion enters or leaves a sphere of radius 7 Å. We can estimate the effect on the potential by a judicious application of Coulomb's law (1). It tells us (see p. 27 of Smythe, 1950) the relation between potential and charge for two concentric spheres, one infinitely far away, which is called the capacitance $C(R)$ of a sphere of radius $R$.

$$\frac{\phi(R)}{Q} = \frac{1}{4\pi\varepsilon_r\varepsilon_0 R} = \frac{1}{C(R)} \tag{5}$$

A single charge $1.6 \times 10^{-19}$ C produces a potential energy of 1 electron-volt in a 7 Å radius sphere, if we use the dielectric constant appropriate for thermal fluctuations at a $10^{-14}$ sec time scale, namely $\varepsilon_r = 2$.

One electron volt is a large fluctuation, compared to the baseline energy of ionic solutions, namely, the thermal energy of $k_B T = 25 \times 10^{-3}$ eV at room temperature. (A fluctuation in potential energy can be neglected only when it is small compared to thermal energy, less than say $0.01 \times 25 \times 10^{-3} = 2.5 \times 10^{-4}$ eV.) Fluctuations of potential energy of some 1 eV—some 4000× this error threshold—occur very often in ionic solutions because it does not take very long for an ion to travel 7 Å: fluctuations of 1 eV occur about twice every picosecond because an ion moving at the speed of sound takes about 460 femtoseconds to travel 7 Å.

If potentials are to be simulated with the precision needed to predict biological phenomenon, somehow these enormous fluctuations and their effects must be correctly averaged. This might be feasible (or anyway safely ignorable) in systems without charge on the boundaries if the effects of fluctuations in potential were linear, e.g., producing a change in concentration proportional to the change in potential. But the coupling between potential and concentration (and other parameters) is very nonlinear, and so correlations may be missed if averaging is only done linearly. Indeed, qualitative phenomena like coupling between movements of ions of different species (Taylor and Krishna, 1993) can be missed if correlations are ignored.

Averaging is made much more difficult because most biological systems contain boundaries with charge. The effects of charge movement are then not a local phenomena. Electric fields in cells extend much larger distances than a Debye length because the membrane of cells acts as a charged boundary. Charge changes potential over a very long distance in cells, unlike in the bulk solution example we just considered (Jack, *et al.*, 1975). "Spherical" cells (i.e., finite cells that are not long like nerve axons) have more or less uniform electrical potential (p. 218 of Kevorkian and Cole, 1996, Barcilon, Cole, and Eisenberg, 1971) meaning that all the ions inside the cell interact with each other and Coulomb's law must be summed over all the ions in the cell. Cells range from a few μm in diameter to say 100 μM. The volumes contain staggering numbers of ions, something like $10^{14}$ if the concentration of ions is 0.15 molar in a cell of 50 μm radius. The electrical potential spreads long distances in nerve axons, typically millimeters. A cylindrical cell 50 μm radius, 2 mm long, contains something like $3\times10^{15}$ ions if it is filled with 150 mM salt solution. Thus, direct calculations of electrical potential in cells are impractical.

It seems intuitively obvious that one should be able to separate the long range and short range effects of the electric field, dealing with one in the average and the other by direct simulation, but this is an unsolved problem, the subject of much on-going research because of the evident consequences of its solution (e.g., Wordelman and Ravaioli, 2000; Keblinski, Eggerbrecht, Wolf, and Phillpot, 2000).

It seems then that direct calculations of biological relevance will not be made with atomic resolution in the near future: neither concentration, nor current can be directly simulated, and electrical potential cannot be simulated when it contains long range components, e.g., when boundary charge is present, as it usually is in biology. Clearly, we must make our goals more modest. We must try to use atomic resolution only where needed, averaging variables everywhere else, thereby keeping the calculation practical and useful for understanding biological systems. Indeed, even if we could do the direct calculation, it would be wise to sacrifice detail and average whenever possible. It is not clear how one would process and understand the results of a direct simulation. The results are enormous numbers of numbers, billions of trajectories, each consisting of billions of numbers. Approximate averaged representations of these trajectories would be needed to understand and design systems, and it seems wasteful to simulate systems directly and then average the results, if one could do an analysis of the averaged system some other way.

***Averaging***. The question is how to do this averaging. The simple answer is that we do not know, nor do we even know if there is a single answer to this question. Different variables and different problems may have to be averaged in quite different ways. This is an essentially mathematical question, with a physical motivation and basis; that is, it is a question in numerical analysis and statistical mechanics. Until those professions provide answers, we must use one description when focussing on atomic structure, and another description when focussing on biologically relevant variables, with no easy way to link the two rigorously.

This chapter presents an averaged macroscopic view of ion permeation, with descriptions and variables appropriate for the biological domain. In particular, we show how to determine the flux (of many ions) through a channel, given a description of its structure, and a description of the driving forces on those ions, namely the gradients of concentration and electrical potential across the membrane. In some simplified cases, we can derive this averaged description from an atomic description, but we certainly cannot do that in general. What is important is that an averaged description of this sort helps understand the behavior of open channels in many conditions of biological and experimental interest.

Let us consider simplified cases first, so we can retain an atomic perspective as long as possible. The customary first description of the motion of atoms since the days of Einstein and Smoluchowski is the high friction version of the Langevin equation (called 'overdamped' in the relevant literature), which describes the thermal (i.e., Brownian) motion of ions in a condensed phase like a solution or protein (Canales and Sese, 1998; Coffey, Kalmykov and Wladron, 1996; Gardiner, 1985; Schuss, 1980). In these solutions, atoms cannot move without colliding with their neighbors: condensed phases contain almost no empty space. Acceleration does not occur on the time scale of interest because of the enormous number of collisions and resulting friction. The same collisions that produce friction inevitably produce noise *Noise$_k$*.

$$\frac{dx_k}{dt} - D_k \frac{e}{k_B T} Z_k E = \textit{Noise}_k = \sqrt{2D_k}\ \dot{w}(t) \tag{6}$$

All the noise *Noise$_k$* in the velocity of each ion is assumed to come from the processes that produce friction. The Langevin equation describes the random motion of a single ion of mass $m_k$ with location $x_k$ at time $t$ subject to an electrical force of $Z_k eE = -Z_k e(\partial \phi / dx)$ where $\phi$ is the electrical potential (in volts) and frictional 'force' with frictional coefficient (per unit mass) $\gamma_k = k_B T / D_k$, where $D_k$ is the diffusion coefficient (units: $m^2/s$), $Z_k$ the valence, $e$ the elementary charge (units: C), $E$ the electric field ($V/m$), and $k_B T$ is the thermal energy (units: joules or $kg\, m^2/s^2$), with $T$ the absolute temperature (units: deg Kelvin) and $k_B$ the Boltzmann constant (units: $J/deg$). Equation (6) has been simplified by 1) considering the high friction limit 2) using the Einstein relation between friction and diffusion coefficients. As written, equation (6) satisfies the fluctuation-dissipation relation.

The Gaussian white noise process arising from the atomic interactions that create friction is $\textit{Noise}_k = \sqrt{2D_k}\, \dot{w}(t)$. Note the dot above the *w*, which denotes differentiation with respect to time. $\dot{w}(t)$ is the infinitely fluctuating white noise familiar to electrical engineers and has dimensions of $t^{-1/2}$. The integral of the white noise is the Brownian motion *w* (without the dot) and has dimensions of $t^{1/2}$. The Brownian motion *w* has the property that if it starts from zero, its average value (indicated by the expected value *E*) is zero, $E(w(t)) = 0$ but its average fluctuation increases indefinitely with time, specifically, $E(w^2(t)) = t$. In particular, the average value of its square is 1 second, at time $t = 1$ s.

It is useful to think of the Langevin equation as a recipe for a simulation because the properties of $\dot{w}(t)$ are bizarre (e.g., $\dot{w}(t)$ crosses and recrosses any line an infinite number of times in any time interval no matter how brief!). That way the mathematical problems involved in defining random motion, which are a real and inescapable consequence of its randomness, are put in a concrete realizable context.

The Langevin equation is a considerable simplification of the molecular dynamics of ion motion (Hynes, 1985; Hynes, 1986). It represents all the interactions of molecules as the result of conservative forces (i.e., that obey conservation of energy, like electric forces or gravitation) and dissipative forces that generate heat at the expense of energy. The randomness of atomic motion is thought to come entirely from the atomic motion that underlies dissipative forces and in its simplest form the Langevin equation is written with the assumption that the electric field and friction are independent of time. These are dramatic simplifications of the real situation and can be expected to be true sometimes and not true others (Rey and Hynes, 1996). What matters here, of course, is whether these simplifications apply to the open ionic channel.

The Langevin equation specifies the motion of one ion, but that is not what we measure when we study current flow through a channel. We measure the movement of many ions and so what we seek is the property of many Langevin equations; we seek to convert the atomic motion of one Langevin equation to an estimate of the variable we measure, current. Fortunately, it is possible to write and solve the equations that describe current (Eisenberg, Klosek and Schuss, 1995) and to simulate them as well, getting the same results (Barcilon, *et al.,* 1993). The result is pleasingly simple and can be written without approximation in the form of a chemical reaction, as we shall soon see.

The flow (i.e., the flux $J_k$ of ion $k$) through the channel is described by the diffusion equation, the Nernst-Planck equation (see Bockris and Reddy, 1970; Newman, 1991).

$$J_k = -D_k(x) A(x) \left( \frac{dC_k(x)}{dx} + \frac{C_k(x)}{RT} \frac{d}{dx} \left[ Z_k F \phi(x) + \mu_k^{ex}(x) \right] \right)$$

$$I = \sum_k I_k = \sum_k Z_k F J_k$$

(7)

$D_k(x)$ is the diffusion coefficient of ion $k$ in the channel. The flux $J_k$ of ions is driven by the gradient of concentration and electrical potential, which together form the electrochemical potential

$$\mu_k = \overbrace{RT\log_e C_k(x) + Z_k F\phi(x)}^{\text{Ideal}} + \overbrace{\mu_k^{ex}}^{\text{Excess}} \qquad (8)$$

Then, we see that flux is proportional to the gradient of electrochemical potential

$$J_k = -\frac{D_k(x)A(x)}{RT}C_k(x)\frac{d\mu_k(x)}{dx} \qquad (9)$$

$\mu_k^{ex}$ is the excess chemical potential produced by such effects as the finite size of ions and charged groups of the channel protein, dehydration/resolvation, and charge transfer through chemical bond formation. The first two terms of equation (8) describe the properties of an ideal gas of charged particles that interact only through their mean electrical potential $\phi(x)$. All other properties are called "Excess". Some of these other properties may arise from physical properties, e.g., the finite size of spherical ions; others might be more chemical, e.g., arising from specific interactions of the atomic orbitals of ion and protein.

Excess chemical potentials can be analyzed (remembering that the excess chemical potential is likely to be a strong function of concentration and other variables) within the traditions of modern electrochemistry, e.g., with density functional theory (Frink and Salinger, 1999; Henderson, 1992; Rosenfeld, 1966; Rosenfeld and Blum, 1986) or the mean spherical approximation of statistical mechanics (Barthel, Krienke and Kunz, 1998; Berry, *et al.,* 2000; Blum, 1975; Blum, Vericat and Degreve, 1999; Durand-Vidal, *et al.,* 1996; Simonin, 1997; Simonin, Blum, and Turq, 1996). We shall show later, briefly, the simplest treatment of these excess chemical potentials, as arising form the finite diameter of ions and charged groups of the channel protein, is enough to account for a range of selectivity phenomena in channels without invoking more chemically specific interactions (although it certainly has *not* been proven that this simple treatment is complete or correct).

The flux can be written (here for the special case where $D_k$ is independent of $x$: the general case is given in the original paper, Eisenberg, *et al.,* 1995).

$$J_k = \overbrace{\underbrace{C_k(L)}_{\substack{\text{Source}\\\text{Concentration}}} \underbrace{\left(\frac{D_k}{d}\right)}_{\substack{\text{Diffusion}\\\text{Velocity}}} \underbrace{\text{Prob}\{R|L\}}_{\substack{\text{Conditional}\\\text{Probability}}}}^{\text{Unidirectional Efflux}} - \overbrace{C_k(R)\left(\frac{D_k}{d}\right)\text{Prob}\{L|R\}}^{\text{Unidirectional Influx}} \qquad (10)$$

$\underset{\text{Channel Length}}{\uparrow}$

In these equations, $d$ is the channel length and the conditional probability $\text{Prob}\{R|L\}$ describes the probability that a trajectory starting on the *L*eft reaches an absorbing boundary on the *R*ight, when a reflecting boundary is placed at the left, just behind the source of the trajectories (i.e., just to the left of the source).

Equation (10) can be written without approximation as a rate equation, namely, 'the law' of mass action

### LAW OF MASS ACTION

$$\boxed{J_k = \overbrace{d \cdot k_f C_k(L)}^{\substack{\text{Unidirectional Efflux}\\ J_{out}}} - \overbrace{d \cdot k_b C_k(R)}^{\substack{\text{Unidirectional Influx}\\ J_{in}}}} \qquad (11)$$

that describes

### PERMEATION AS A CHEMICAL REACTION

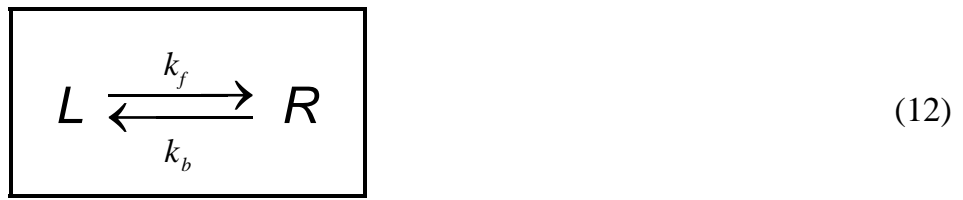

$$\boxed{L \underset{k_b}{\overset{k_f}{\rightleftarrows}} R} \qquad (12)$$

In these equations, the rate constants $k_f$ and $k_b$ are

$$k_f \equiv \frac{J_{out}}{C_k(L)} = k\{R|L\} = \frac{D_k}{d^2}\text{Prob}\{R|L\}; \quad k_b \equiv \frac{J_{in}}{C_k(R)} = k\{L|R\} = \frac{D_k}{d^2}\text{Prob}\{L|R\} \qquad (13)$$

It is important to keep in mind that the law of mass action is not a general physical law. Rather, it needs to be derived from the physical laws underlying a system and will in general be a good approximation in some conditions, but not in others. Eisenberg, *et al.,* 1995, can be viewed as a stochastic derivation of the law of mass action (eq. (12)) that precisely defines the conditional probabilities of eq. (13) and shows how they can be computed or simulated. In this way, 'the law of mass action' is shown to be valid provided bath concentrations and *trans*membrane potential are maintained fixed, and $\phi(x)$ does not vary as the concentrations $C_k(L)$ or $C_k(R)$ or *trans*membrane potential are varied. Surprisingly, if the law is valid at all, the derivation shows *it is valid for any shape of the potential barrier.* In this way, the metaphor of channel permeation as a chemical reaction (Eisenberg, 1990) can be made exact. Indeed it *is* exact when the *trans*membrane potential and bath concentrations are kept fixed in a voltage clamp experiment.

The rate constants and conditional probabilities can be written particularly neatly when friction is large and simple in behavior, described by a single diffusion coefficient, a single number $D_k$ for each species *k* of ion. Then, if the potential across the channel is $V_{appl}$, namely $\phi(L) - \phi(R)$

$$k_f = k\{R|L\} = \frac{D_k}{d^2}\text{Prob}\{R|L\} = \frac{D_k}{d^2} \cdot \frac{\exp(Z_k FV_{appl}/RT)}{\frac{1}{d}\int_0^d \exp(Z_k F\phi(\zeta)/RT)d\zeta}$$

$$k_b = k\{L|R\} = \frac{D_k}{d^2}\text{Prob}\{L|R\} = \frac{D_k}{d^2} \cdot \exp(Z_k FV_{appl}/RT)\frac{1}{\frac{1}{d}\int_0^d \exp(Z_k F\phi(\zeta)/RT)d\zeta} \qquad (14)$$

and the flux and current through the channel is

$$J_k = \overbrace{D_k \frac{C_k(L)\exp(Z_k FV_{appl}/RT)}{\int_0^d \exp(Z_k F\phi(\zeta)/RT)d\zeta}}^{\textbf{Unidirectional Efflux}} - \overbrace{D_k \frac{C_k(R)}{\int_0^d \exp(Z_k F\phi(\zeta)/RT)d\zeta}}^{\textbf{Unidirectional Influx}} \quad ;$$

(15)

$$I = \sum_k I_k = \sum_k Z_k F J_k$$

At first, it may seem redundant to derive eq. (15) because eq. (7) – (13) give the flux as well. Eq. (15) is different, however, because it gives the flux in terms of the variables controlled experimentally, the bath concentrations $C_k(L)$, $C_k(R)$ and the membrane potential $V_{appl} = \phi(L) - \phi(R)$. In this case, we can also write explicit expressions for the concentration everywhere, namely

$$C_k(x) =$$

$$\frac{C_k(L)\cdot \exp\big(Z_k F[V_{appl} - \phi(x)]/RT\big)\cdot \int_x^d \exp(Z_k F\phi(\zeta)/RT)d\zeta}{\int_0^d \exp(Z_k F\phi(\zeta)/RT)d\zeta}$$

$$+ \frac{C_k(R)\cdot \exp(-Z_k F\phi(x)/RT)\cdot \int_0^x \exp(Z_k F\phi(\zeta)/RT)d\zeta}{\int_0^d \exp(Z_k F\phi(\zeta)/RT)d\zeta}$$

(16)

remembering that $C_k(x)$ describes the probability of location of an ion. These expressions can be easily generalized if $D_k$ depends on location (Nonner, Chen and Eisenberg, 1998; Nonner and Eisenberg, 1998).

It may seem that we have solved the problem and created a macroscopic description of the flux through a channel that includes some atomic detail. But this is **not** the case because we have not completely specified how the permeating ions interact with the channel protein. The analysis up to here describes only the friction ions experience as they flow through a channel as do all diffusional theories of permeation, starting with the constant field theory of Goldman, 1943, and Hodgkin and Katz, 1949. But the analysis up to here does not describe how the ion interacts with the electric charge on the channel protein because it does not describe how the electrical potential $\phi(x)$ is produced. An explicit expression for the potential cannot be written because the Poisson equation defining it (eq. (17), below) has a second derivative in it, whereas the diffusion equation (7) defining the flux and concentration only contains first derivatives.

**<u>*Charge is the source of potential*</u>**. It is important to remember that the fundamental source of the electric field is charge; charge is the fundamental electrical property of matter; potential is the outcome of the charge. Potential is the outcome of net charge. All matter contains charge; when its negative and positive charges are separated in space, a potential is created. This potential cannot be maintained at a constant value if charges move, unless some active process compensates for the effect of the charge movement.

When a potential is maintained at a nonzero value, for example, in the wires that bring electricity into our homes, or in the output of a battery or power supply, it is done so by a machine that separates and supplies charge, a generator, or a chemical reaction, or a circuit of transistors, in these examples. These machines use energy to separate charge. Proteins and channels have limited access to energy; if they do not 'burn' ATP, or interact with pH buffers, they cannot change their charge. Thus, *a channel protein must be described as distribution of charge, not of potential*. The potential profile in a protein cannot be held constant. The potential distribution within a channel is necessarily a variable but the potential in the baths can be maintained constant by a voltage clamp amplifier (that uses energy to separate charge). Experimental apparatus cannot maintain a profile of potential within a channel because it cannot supply charge and energy inside the channel. Even if apparatus could maintain a potential at one location in a channel, it would significantly disturb the potential at another location. Charge added to a location inside a channel (e.g., by a mutation) is nearly certain to change the potential profile everywhere in the channel and usually in a significant way (since the current we measure experimentally is an exponential function of the potential profile, speaking loosely). Interestingly, at one location a potential can be maintained constant at a zero value by a passive device, simply a wire connected to ground. In that case, the energy to move charge through the wire comes from the electric field itself, that is generated by the separation of charge between system and boundary.

Rate constants that determine flow through a channel (or chemical reaction) depend on the potential profile; they do not depend directly on the charge in the channel protein or surrounding baths as eq. (14) makes clear. The rate constant of the law of mass action necessarily depends on the potential profile that limits and creates the rate of a chemical reaction. The potential is determined indirectly by the charge and the structure of the system and thus so is the rate constant. *The rate constant cannot be expected to be constant*, because changes in the potential in the bath, changes in the composition of the bath, changes in the charge or structure of the channel protein (produced by mutations, for example), will change the potential profile and thus the rate constant. We must determine the potential profile before we can determine the flux through a channel.

**_From charge to potential_**. In the real channel, charges are moving rapidly and incessantly, and so the potential is changing on the time scale of atomic motion, namely $10^{-16}$ sec. The currents of biological and experimental interest are on the $10^{-5}$ to $10^{-3}$ sec time scale. Clearly, averaging is needed to determine the potential profile relevant to permeation. The question is how to do that averaging.

In general, theories must average in a way that does not mix disparate quantities. For example, when studying the respiratory system it is reasonable to study breathing in the average lungs of all humans. But when studying the reproductive system, it is not reasonable to average over all human beings; rather, one must separate humans into two groups, the functionally relevant classes, and then average over each of those. Similarly, when studying single filing, one must study ions that flow left to right, that flow right to left, that flow left to left, and that flow right to right. Each must be averaged separately; the potential profile must be computed for each; and the size of each class of trajectories must be determined (e.g., what fraction of trajectories are *cis LL* and so on). This task has only begun (Barkai, Eisenberg and Schuss, 1996). Here we use the simplest possible mean field approximation, saying that the charge averaged over the time of biological interest creates the potential of biological interest. Interactions in charge movement (e.g., single filing phenomena) are ignored. The resulting theory must not be used to describe single file phenomena, but it can be used to describe net fluxes, and the electrical currents they produce. We present the one dimensional form of the theory here since that is most applicable to the great majority of channels, where structure is not known. When structure is known in atomic detail, a three dimensional version of the theory should be used (Hollerbach*, et al.,* 2000; Kurnikova*, et al.,* 1999). The validity and utility of the theory is shown by the range of conditions in which it can fit experimental data using parameters of definite meaning and reasonable value, and hopefully not too many of them.

Once averaging has been defined, the potential must be determined by Coulomb's law or the Poisson differential equation, which is its exact equivalent: those are the laws of electrostatics. If other forces are involved, they will contribute to the potential energy as well, but again how they contribute must be determined from the fundamental properties of those forces, usually described by a set of differential equations that need to be satisfied along with those already stated, e.g., Chen*, et al.,* 1995.

Here we start with the radical working hypothesis that **only** electrostatic forces are involved in permeation. We suppose that all interactions of the channel protein and permeating ion are described by Poisson's equation, along with the frictional constant of the Nernst Planck equations. Obviously, this is just a starting place, adopted for the sake of simplicity. Chemical forces will be added later, as needed, but the philosophy of this approach is to do the electrostatics carefully and add additional chemical forces only where experiments show they are needed.

Electrostatics dominate the problem because of the large charge densities and tiny volume of channels. The tiny volume of a channel (a 10 Å long channel of 4Å diameter has a volume of $5\times10^{-28}$ m$^3$) implies that even one ion produces a large concentration. If the 1 nm long channel considered above had a total permanent charge of –1 (e.g., it had one acidic amino acid residue), it would contain about one mobile charge of opposite sign to maintain approximate electrical neutrality, and the concentration of mobile charge would be around 13 Molar! This is an extremely highly concentrated solution, nearly a solid, with about 25% of the molecules being ions and a roughly similar fraction of the volume being occupied by ions. The properties of such solutions are dominated by their electrostatics, along with chemical contributions produced by the finite volume of the ions.

Now, we treat just the electrostatics, using the one dimensional version of Poisson's equation for the sake of simplicity, but clearly the three dimensional version should be used wherever it can be (Kurnikova, *et al.,* 1999; Hollerbach*, et al.,* 2000). We define the mean electrostatic potential for all ions as the solution $\phi(x)$ of the differential equation

$$\varepsilon_0 \left[ \varepsilon_r(x)\frac{d^2\phi}{dx^2} + \left(\frac{d\varepsilon(x)}{dx} + \varepsilon_r(x)\frac{d}{dx}\left[\log_e A(x)\right]\right)\frac{d\phi}{dx} \right] = -\rho(x) \qquad (17)$$

where the average charge $\rho(x)$ does not include the induced charge. $\varepsilon_0\varepsilon_r(x)$ is the permittivity $\varepsilon_0$ of free space times the relative (dimensionless) dielectric coefficient $\varepsilon_r(x)$; $A(x)$ is the cross sectional area of the channel. The charges that create the potential are

$$\rho(x) \equiv eN_A \left[ P(x) + \sum_k Z_k C_k(x) \right] \qquad (18)$$

The charge $\rho(x)$ consists of

(1) the charge $eN_A \sum_k Z_k C_k(x)$ of the ions (that can diffuse) in the channel, of species $k$ of charge $Z_k$, and mean concentration $C_k(x)$; $N_A$ is Avogadro's number; typically $k = \text{Na}^+, \text{K}^+, \text{Ca}^{++}$, or $\text{Cl}^-$ and

(2) the permanent charge of the protein $P(x)$ (mol≅m$^{-1}$), which is a permanent part of the atoms of the channel protein (i.e., independent of the strength of the electric field at $x$) and does not depend on the concentration of ions, etc, and so is often called the fixed charge. $P(x)$ is really quite large ($\sim 0.1$ to $1e$ per atom) for many of the atoms of a protein and wall of a channel.

(3) The dielectric charge (i.e., the induced charge which is strictly proportional to the local electric field) is not included in $\rho(x)$ because it is described by $\varepsilon(x)$. It is generally small compared to the structural charge and is neglected in our discussion (but not in our software). In this primitive model of ion permeation, hydration and solvation forces are only represented by their dielectric effects.

We are now nearly finished with the physical description of our problem at its simplest level. The transport equation (7) and the Poisson equation (17) together specify the potential and concentration of ions everywhere in the channel, when they are supplemented by boundary conditions that describe how potential and concentration are controlled in the baths outside the channel. Once these Poisson-Nernst-Planck equations are solved, substitution into equation (15) & (7) yields the current observed. These equations are sometimes called *PNP* for short and for my amusement because they are nearly identical to the drift diffusion equations of semiconductors, including *PNP* transistors, see Lundstrom, 1992; Selberherr, 1984; Sze, 1981.

It is important to realize that the *PNP* system of equations is a complete theory. Once the properties of the channel and permeating ions are specified, once the *trans*membrane potential and concentrations in the baths are specified, all currents of all ions are predicted at every potential, simply by solving the equations. In particular, the channel is specified by its geometry and its distribution of fixed charge. Its interactions with the permeating ions come from the friction experienced by the ion (described here by a diffusion coefficient for each ion in the channel, which is of course different from the diffusion coefficient for the ion in the baths, since the local environment of the ion is so different in the channel and bath); from the excess chemical potential $\mu_k^{ex}$ if it is present; and from the electrostatic and diffusional forces. Those forces are outputs of the theory and are not adjustable independently of each other. That is why the theory is called self-consistent: all the variables *together* must satisfy the *PNP* equations and boundary conditions. Of course, the predictions may be wrong, the channel may do more to the ions than implied by the equations, but that is the point of experimentation, and if the experimentation proves the theory wrong, then one adds more physics and chemistry to the theory, or chooses a different approach altogether.

**<u>Solving the PNP equations</u>**. Before we can use the *PNP* equations, we must solve them. The equations (17) & (7) are rewritten here in their simplest form for a channel of constant diameter, with no excess chemical potential, and dielectric and diffusion coefficient independent of location

$$\varepsilon_0 \varepsilon_r \frac{d^2\phi}{dx^2} = -eN_A \left[ P(x) + \sum_k Z_k C_k(x) \right] \tag{19}$$

$$J_k = -AD_k \left( \frac{dC_k(x)}{dx} + \frac{z_k F}{RT} C_k(x) \frac{d\phi(x)}{dx} \right)$$

$$= -\frac{AD_k}{RT} C_k(x) \frac{d\mu_k(x)}{dx}$$

(20)

We must solve these equations for the concentration $C_k(x)$ and potential profiles $\phi(x)$ within the channel, given the concentration of ions $C_k(L)$ & $C_k(R)$ in the baths, the transmembrane potential $V_{appl}$, the distribution of fixed charge $P(x)$ along the channel protein, and the value of the diffusion coefficient and dielectric coefficient within the channel (Fig. 1). Then, we take those profiles and use them to compute the current through the channel, from eq. (15) & (7). Solving these equations in one dimension is an easy task because software is available to do it. A program is available by anonymous FTP from ftp.rush.edu (in /user/Eisenberg/Hollerbach) that implements a back and forth iteration scheme that is very fast and accurate (to more than 15 significant figures in some 20 iterations that typically take less than 0.1 sec to compute). Extensive experience and mathematical and numerical analysis in the semiconductor community shows that only the back and forth scheme originally introduced by Gummel yields solutions of tolerable accuracy (Jerome, 1995; Selberherr, 1984). Typical integration schemes depending on discretization and matrix inversion will not work for equations of this sort.

It would be most helpful at this stage to give analytical formulas for simple cases. Unfortunately, these are not known, although much effort is being made to derive them. We must depend then on numerical solutions.

***What do we do with these equations***? The numerical solution of the PNP equations give a complete description of the channel in a certain sense; but how do we use that solution to understand biology?

Fig. 1

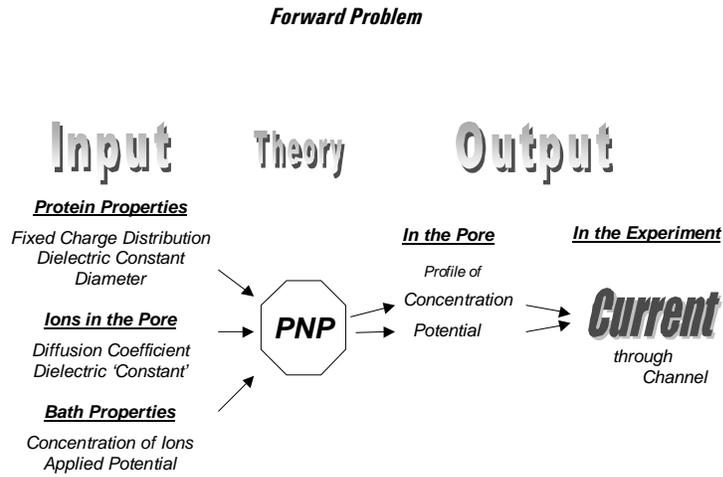

Fig. 1. Forward Problem. The inputs and theory needed to predict the current through the channel.

What we do depends on our knowledge of structure. If we do not know the structure, we must do the inverse problem shown in Fig. 2 and discussed later to determine the profile of effective fixed charge $P(x)$. That profile can then be used to predict $I(V)$ curves in conditions different from those used to determine $P(x)$ curves. Comparison with reality is then a check on the theory.

If we do know the structure of the channel, we can convert the structure into a prediction of current voltage relations ($I(V)$ curves) as shown in Fig. 1. We take the known structure, and assign charges to each of the atoms of the protein, using one of the well distributed programs of molecular dynamics. MOIL is one such program and is available free from web site www.tc.cornell.edu/reports/NIH/resource/CompBiologyTools/ maintained by its author Ron Elber. While there are ambiguities and uncertainties in the choice of charges, these have only second order effects on predictions of $I(V)$ relations. In these calculations, the protein has a shape as well as a distribution of fixed charge. This is represented by a 'diameter' in a one dimensional calculation. The diameter varies with location in the protein and that variation is included in the theory. The shape of the protein is rarely known at all because of the difficulties of crystallization and crystallography. It is almost never known in a range of solutions, so the assumption is made that the structure of the protein does not change as transmembrane potential changes or the type and concentration of salt is changed in the baths. This assumption is certainly violated in some cases, and represents a serious restriction on the validity of any theory.

Similar structural assumptions are commonly made when a protein's function is studied by making mutations. Here the assumption of no structural change is particularly troublesome because crystallography of porin (e.g., Jeanteur, *et al.,* 1994; Saint, *et al.,* 1996; Schirmer, 1998) shows that most mutations produce significant structural changes. Studying mutations without knowing structure is likely to be frustrating for this reason. In the worst case, each mutation has to be treated as (nearly) a new protein of unknown structure.

Fig. 2

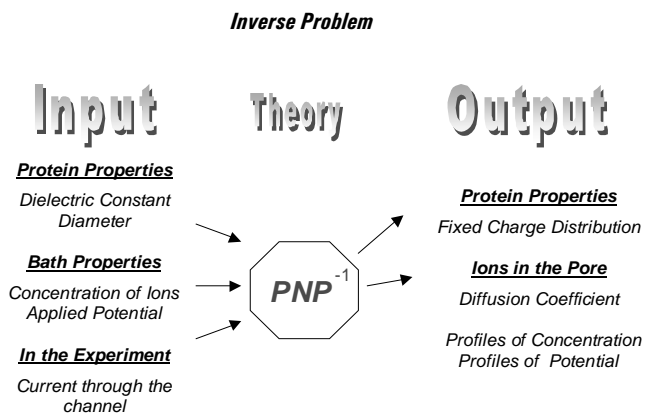

Fig. 2. Inverse Problem. The assumptions and experimental results which are inputs to the theory, used to determine the properties of the channel protein. Once the protein properties are determined, they can be used in the Forward Problem of Fig. 1 to predict properties of channels in other situations.

The calculations of function from structure are really calculations of averaged function (over the biological time scale) from average structure. The thermal fluctuations in structure are dealt with implicitly, as are the fluctuations correlated with ion movement in the channel. The effects of fluctuations are included in the diffusion coefficient and dielectric constant (and fixed charge density and diameter) of the theory, although how well they are included is an open question.

***Diffusion coefficient***. One effect of the protein on the ion is represented by the diffusion coefficient of each ion within the channel. The diffusion coefficient represents the friction on the ion, that is to say, the force opposing ionic motion proportional to the velocity of motion of the ion. This friction arises from collisions between the ion and adjacent atoms of water and the protein. Movement in condensed phases always involves collisions becomes condensed phases like proteins and water contain essentially no empty space.

Friction also arises from electrostatic interactions of the ion with surrounding charges. These are forced to move when the permeating ion moves, because the electric field is so strong. When these neighboring charges move, they collide, dissipating mechanical and electrical energy into heat. (To be precise, collisions change a translational component of the distribution of velocities into a component with no preferred direction; that is to say they convert translational motion into the random motion we call heat: Brush, 1986.) The permeating ion in the channel provides the energy that is converted into heat and so experiences what is called dielectric friction.

Dielectric friction (as it is called) is thought to be responsible for most of the friction in a bulk solution or channel.

Both collisional and dielectric friction should in principle be different at different locations in the channel. In our formulas, the diffusion coefficient is allowed to depend on location, but we do not have any good way to know this dependence and thus often assign it a value independent of location. Of course, each ion in each type of channel has a different diffusion coefficient (in general), but the diffusion coefficient assigned to an ion in a particular type of channel is **not** adjusted as concentration or membrane potential changes. The fits of *PNP* are achieved with single values of the diffusion coefficient for each ion in each channel type at all potentials and concentrations. It is remarkable that a single value of the diffusion constant is able to fit $I(V)$ relations measured in a wide range of solutions. While this result might be coincidence, reflecting the limited number of channels and solutions studied to date, it seems to me more likely to be meaningful, perhaps a result of the nearly solid environment in the highly concentrated interior of the selectivity filter of channels.

The channel's interaction with the ion is also governed by the dielectric constant. The dielectric constant measures the amount of charge induced by the presence of a fixed charge. Because matter is held together by electrical forces produced by only a handful of charges in a given molecule, it is not possible to introduce a charge into a channel without distorting the charge distribution of the channel protein. A helpful analogy is with the gravitational interaction of two nearby binary stars. In that case, the shape of each star (i.e., its distribution of matter) is affected by the other star; one cannot neglect the effect of one on the other, as one can neglect the effect of your body mass on the shape of the earth, when measuring your own weight.

The distortion in charge in the protein is produced by many complex processes that take a long time to develop after an ion is moved; a few nanoseconds is a reasonable estimate. All these processes are crudely lumped into a dielectric constant in *PNP*; indeed, the dielectric constant also describes the process of solvation, namely the interaction of the ion with the channel wall and water. This crudity is a serious limitation to *PNP*, but not as serious as it seems: direct calculations show that varying the dielectric constant has surprisingly little effect on predicted $I(V)$ relations, presumably because of the effects of fixed charge.

The high density of fixed charge in channels seems to be a key to understanding permeation. As we have discussed, the fixed charge on the wall of a channel requires the presence of nearby mobile charges. These ions are found (mostly) in the pore of the channel, which is tiny, and so the resulting ionic concentrations are many molar. In solutions of this concentration, with this much fixed charge present, one is in a very special domain, evidently designed by biology for its purposes. In this domain, direct calculation shows the dielectric constant has only second order effects on predictions of $I(V)$ relations. General considerations (Henderson, Blum and Lebowitz, 1979) show that correlation terms (like induced charge) are much smaller than mean field terms when fixed charge density is high. Later we shall see how the dehydration resolvation energies accompanying ion movement can be included in *PNP* by introducing an excess chemical potential.

**_Driving Force for ion movement_**. The movement of ions through the channel are also determined by the driving forces on the ion, namely the concentration and electrical potential which determine the electrochemical potential of the ion, its free energy per mole. These are inputs to the *PNP* calculation, and are known and determined by the experimenter. Of course, setting the electrical potential is hardly a trivial matter. In experiments, an elaborate apparatus, the voltage clamp, is used to separate the charge (between the baths on either side of the channel) and thereby keep the potential difference between them constant. The *trans*membrane potential can be maintained constant only over a limited bandwidth, typically some 10 kHz, because at higher frequencies more and more charge must be supplied per unit time (i.e., more current) to maintain the electric field. In the language of engineering, a large amount of current is needed to charge the capacitances of the system and the voltage clamp amplifier has limited ability to supply charge. Even if potential is maintained at zero by a direct connection to ground, a simple calculation shows that potentials cannot be maintained constant on atomic time scales. The currents needed are enormous and cannot be supported by wires and ordinary apparatus (indeed, the currents necessary to control potential on a femtosecond time scale, or even a picosecond time scale, would fall far outside the domain of electrostatics and would in fact radiate substantial energy as radio waves). Fortunately, these problems are solvable and solved *on the biological and experimental time scale* $>0.1$ msec by apparatus that is commercially available. Voltage clamping is a practical even routine procedure on the biological but not atomic time scale.

In the animal, a voltage clamp is not available, and cells fall broadly into two classes, those in which the voltage is maintained more or less constant (called 'inexcitable' cells) and those in which it changes quickly and are called 'excitable' cells. We will not discuss how to deal with the latter situation, which was the main topic of physiological investigation from Volta and Galvani to Hodgkin (who essentially solved the problem). The book by Weiss (Weiss, 1996) provides excellent introductions to this classical work.

Control of concentration is both easier and harder than control of potential. It is easier in the case of most ions, because they are present in substantial concentrations on both sides of the channel. In that case, the number of ions that flow through one channel does not significantly perturb the concentration of ions near the channel, and the concentration is naturally controlled. Of course, this situation is not absolute: the flow of ions is accompanied by a resistive potential drop (called an IR drop in electrical engineering, in celebration of Ohm's law) that is not always negligible. Furthermore, when ion concentrations are tiny, as they are inside cells for $Ca^{++}$, concentrations are not necessarily constant and the variation of concentration must be included in the theoretical description. One way to do that, which is mathematically convenient, is simply to extend *PNP* into the baths, since it describes changes in concentration in response to current flow, but this is not enough if specialized binding systems or structures significantly modify $Ca^{++}$ concentration, as is usually the case. Here we only treat the simpler situation.

This ends the catalog of variables of the *PNP* equations. Given the parameters shown in the left side of Fig. 1, or determined by the inverse process shown in Fig. 2, the *PNP* equations predict the current through the channel. Rather, they predict the profile of potential and concentration in the channel eq. (17) & (16), from which the current can be predicted by simple integration, as we have seen eq. (15). The theory can be checked by varying potential and concentration in the bath and seeing if the current varies as predicted. Usually the diffusion coefficient and dielectric constants are not known, even if the structure is, so an absolutely direct or forward calculation is not possible. Usually, one assigns a diffusion coefficient and dielectric constant so one $I(V)$ relation is fit and then sees how well those parameters explain the properties of the channel in general, that is to say, one sees how well a single value of diffusion and dielectric coefficient can explain the $I(V)$ relations measured in a range of solutions.

**<u>*Biology of permeation*</u>**. If the theory fits a reasonable range of data, one can begin to do the more usual biology of permeation. One can then ask how the $I(V)$ relation is produced? How does it change as ion types change? How does a change in the protein's charge or shape change the biological function of the molecule?

These questions are answered by directly calculating the profile of electrical potential in the channel and the profile of concentration of each ion. Looking at these, one can develop insight into how the $I(V)$ relation is produced by the shape and charge of the protein, and the concentration and potentials of the bath.

A general understanding of the role of these variables is not yet available, although it is being worked on. Certain principles have already emerged. If the channel has more or less uniform fixed charge (i.e., it has no regions where the charge changes sign or reaches zero), the action of the fixed charge is to *buffer the concentration of ions in the channel*. Electroneutrality guarantees that the number of ions in the channel nearly balances ('neutralizes') the fixed charge on the wall of the channel, over a wide range of applied potentials and concentrations of ions. The *occupancy of the channel is determined by the charge on the walls of the channel*, to a large extent. Thus, the contents of the channel are quite constant, and the resistance/conductance of those contents are quite constant as well. What variation that does occur is at the edges of the channel, where boundary layers do vary with concentration and potential. Such boundary layers are not without importance, because they are in series with the channel, and they in fact regulate the properties of analogous systems in semiconductors (Streetman, 1972) and make them into useful devices like transistors. Nonetheless, it is customary to ignore the boundary layers in a first order qualitative analysis of channels. (Boundary layer potentials are enough to explain many of the anomalous properties of ion channels, e.g., Nonner, *et al.,* 1998; Nonner and Eisenberg, 1998.)

**_Linearity of I(V) relations_**. In general *PNP* predicts single channel $I(V)$ relations that are fairly linear, even when the concentrations of ions are very different on the two sides of the channel because of the buffering effect just discussed (see Fig. 6 & 7). Or to put it more precisely, *PNP* predicts that $I(V)$ relations will not change much, as solutions are changed on one side or the other of the channel. This behavior of *PNP* is strikingly different from the behavior of constant field or barrier models, which invariably (in the first case) or usually (in the second case) show large changes in $I(V)$ relations in asymmetrical solutions, with different concentration of ions on both sides. While little data of this sort has been published, what has been published is quantitatively in agreement with *PNP*. Anecdotally reported experiments suggest that many other channels have $I(V)$ relations that are surprisingly insensitive to asymmetry in bath concentrations. If asymmetry in bath concentrations is found to change $I(V)$ relations, *PNP* may, or may not be able to fit the data, depending on the many factors, particularly the properties of boundary layers and anomalous properties produced by localized excess potentials (Nonner, *et al.,* 1998).

**_What if we do not know the structure_**? In most cases, the structure of the channel is not known, and we must proceed with an inverse problem, as shown in Fig. 2. Here we must determine the profile of fixed charge $P(x)$ that produces the observed $I(V)$ relations, along with the diffusion coefficient of each ion, and the dielectric constant. Once $P(x)$ is determined from data in one set of solutions, it can be used to predict $I(V)$ relations in other solutions. Comparison with reality serves to check the theory.

This inverse problem can be solved with reasonable reliability if one works hard and is clever. If experiments are available in many solutions, it has been possible to determine the effective fixed charge density. If mutations in the channel are available—that change the charge, for example—it is possible to see if they fit as well, assuming that the protein structure is not changed by the mutation. In all these calculations, it is best to treat the diffusion coefficient as a constant, independent of location, with a different (but constant) value for each type of ion.

Much can be learned about permeation this way, because the profile of potential and concentration in the channel can be calculated for each applied potential, concentration, and mutation of interest. One develops molecular and atomic insight by varying the charge on the protein (or permeating ion), by varying its diffusion coefficient, for example, and seeing how that changes the profiles of potential and concentration of each ion, and thereby the $I(V)$ relation of the channel.

The fixed charge profile $P(x)$ determined this way is an effective profile and its relation to the charge density of the actual protein is not yet known. This is a complex and subtle problem even without the ambiguities of curve fitting (i.e., we do not know how to derive the effective one dimensional profile from a known three dimensional profile) and requires much more attention.

***What are the general conclusions***? *PNP* is a young enough theory that many general conclusions are yet to be discovered. One conclusion is overwhelmingly clear, however, and that is the importance of the potential profile (Eisenberg, 1996). *PNP* shows what is clear on very general grounds: the profile of potential in a channel is a sensitive function of ionic concentration in the bath, and of the applied potential, the charge on the protein, and most other parameters of the system.

The potential profile is not constant in any sense. It varies in space and it varies with experimental conditions. The potential profile is produced by the fixed charge on the protein and the charges applied to the boundaries to maintain the applied potential; the charges in the bath rearrange themselves to balance the fixed charge of the protein and the applied charge on the electrodes. The rearrangement of charge reduces the effects of the charge of the protein and the electrodes; the bath ions are said to shield those charges.

Shielding phenomena of this sort play a dominant role in many electrical phenomena, for example, the effects of a Faraday cage of metal on the potential inside, and it should not be surprising that they are dominant in channels.

We now look at the size of these effects for a particular channel gramicidin. Here Uwe Hollerbach has calculated the potential distribution inside a gramicidin with the charge distribution shown in Fig. 3. (The calculation was done in three dimensions but the averaged charge is shown for reasons of visual clarity. The distribution of permanent charge is not symmetrical because an energy optimized structure was used that happened not to be symmetrical, see Elber, *et al.,* 1995). Fig. 4 & 5 show that the potential profile is a sensitive function of concentration. Note the scale of the ordinate. We have commented that potential energy changes must be small compared to $k_BT \sim 25$ meV if they are to be negligible. The shielding effects seen here are not small. Fig. 6 & 7 show the corresponding $I(V)$ curves and Fig. 8 shows the occupancy. Note that the occupancy is not particularly well buffered in gramicidin because gramicidin has zero net charge. In channels with net charge on their wall, occupancy is buffered more strongly.

The existence of shielding has profound effects on our qualitative understanding of channels. Eq. (12) through (15) show that the flux and current through a channel depend **exponentially** on the potential profile. In particular, the rate constant depends **exponentially** on the potential profile. Fig. 4 & 5 show that the potential profile is a sensitive function of concentration and we have argued (with some rhetorical excess) that the potential profile is a sensitive function of everything, hardly ever constant in space or with changes in conditions. Thus, **rate**

Fig. 3

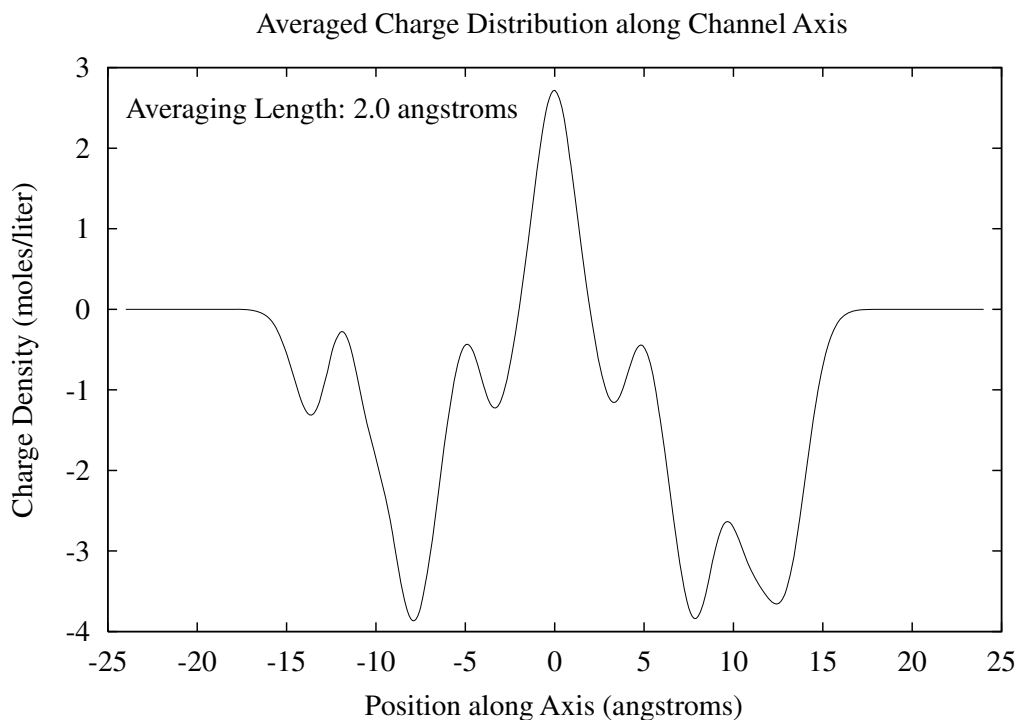

Fig. 3. The averaged charge density along gramicidin. Calculations were done in three dimensions but the averaged charge is shown for reasons of visual clarity. The distribution of permanent charge is not symmetrical because an energy optimized structure was used that happened not to be symmetrical, see Elber, *et al.,* 1995.

Fig. 4

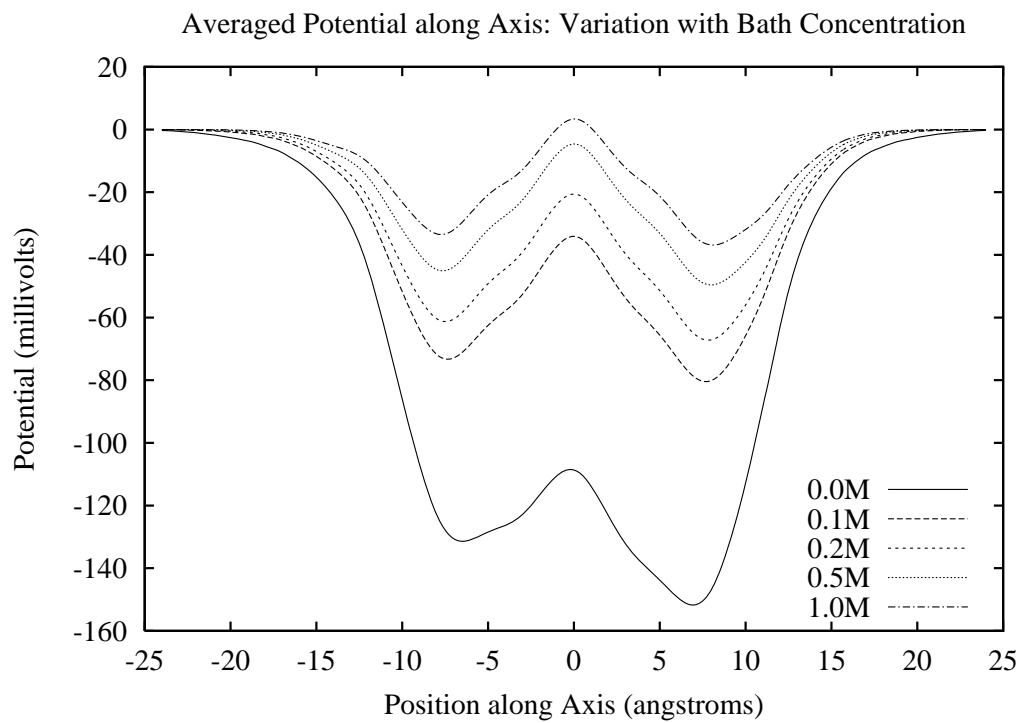

Fig. 4. Averaged potential along the channel at different bath concentrations. Note the large effect of salt concentration on the potential profile. Such effects are caused by the different amounts of screening (i.e., shielding) of the fixed charge on the channel protein.

Fig. 5

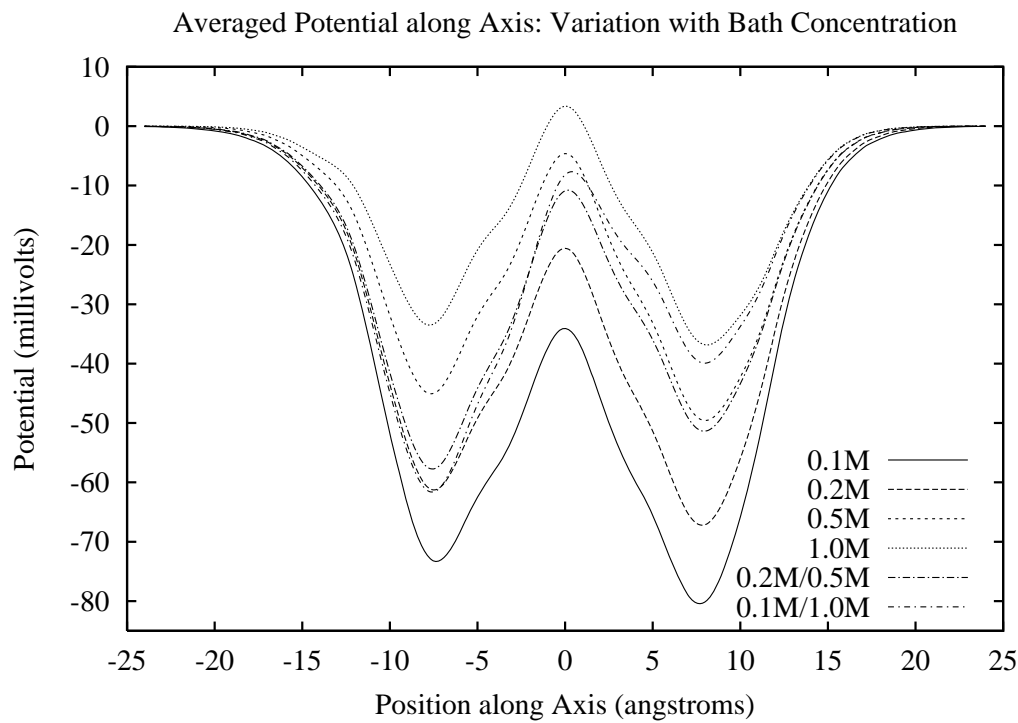

Fig. 5. Averaged potential along the channel at different bath concentrations. Note the large effect of salt concentration on the potential profile. Such effects are caused by the different amounts of screening (i.e., shielding) of the fixed charge on the channel protein.

Fig. 6

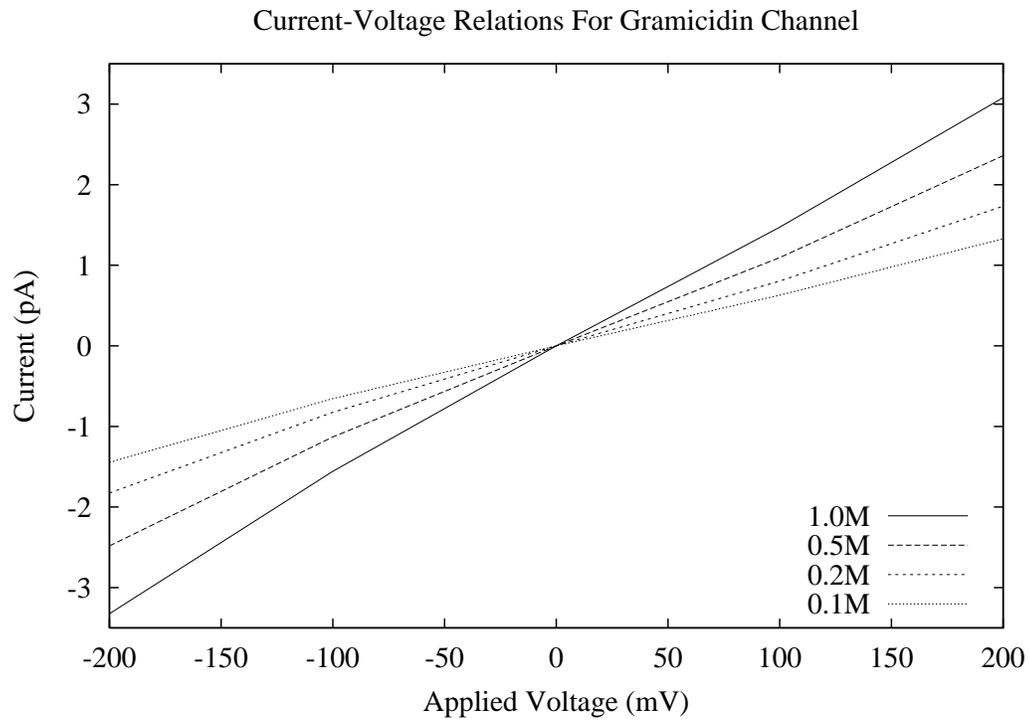

Fig. 6. Current voltage relations in the solutions shown in Fig. 4 and 5.

Fig. 7

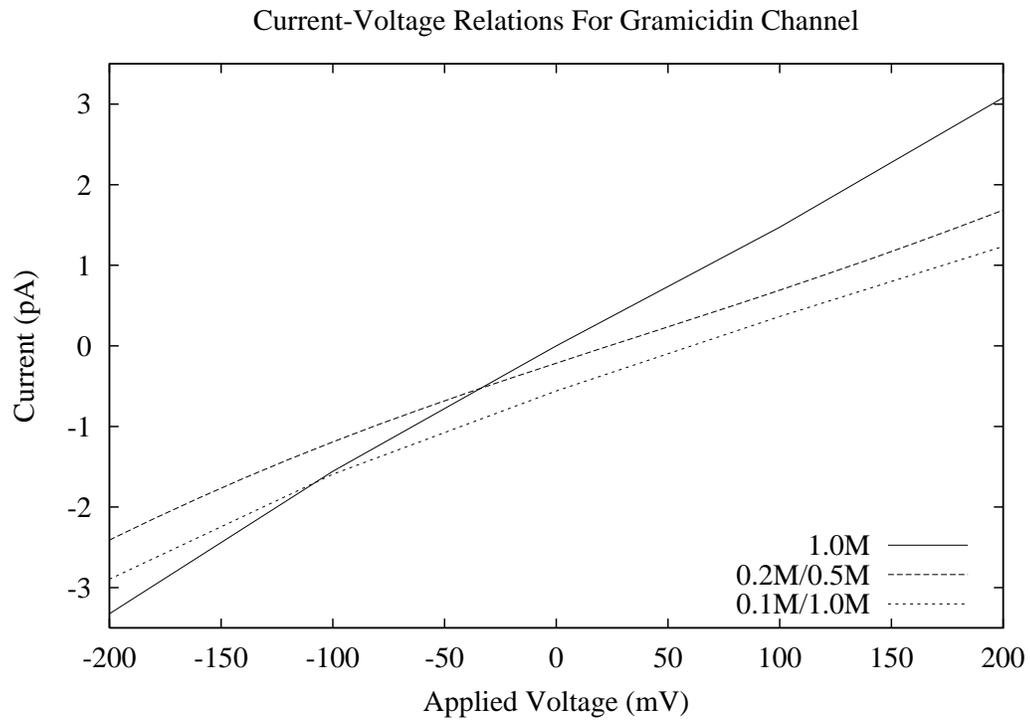

Fig. 7. Current voltage relations in the solutions shown in Fig. 4 and 5.

Fig. 8

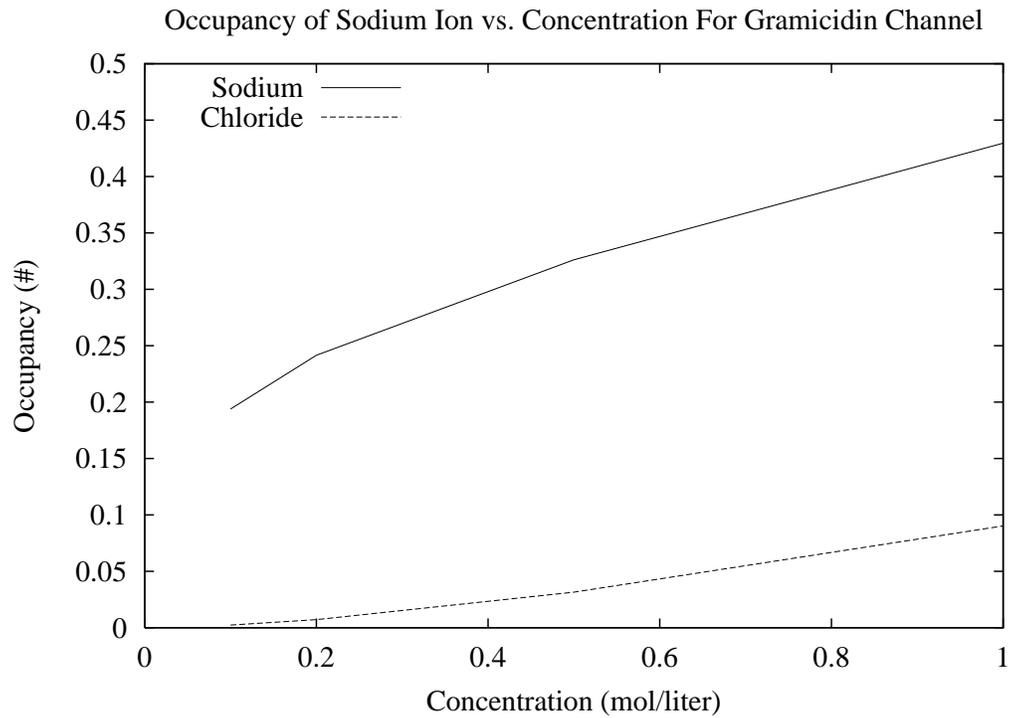

Fig. 8. The occupancy (i.e., integral of concentration) of the channel in different solutions when the *trans*membrane potential is zero.

***constants must depend a great deal on concentration*** and other conditions. Since rate constants are almost always assumed independent of concentration in 'Eyring style' models, there is evident difficulty here. This difficulty extends to the study of proteins and active sites in general (Eisenberg, 1990), where rate constants are nearly always assumed independent of concentration and other parameters that are certain to change shielding and potential profiles, in the majority of cases.

The dependence of potential profiles on concentration and shielding shown in Fig. 4 & 5 is clear, but what is not clear are the general implications. These are best seen in a related field of science, computational electronics, where equations very much like *PNP* are used to describe the current flow through diodes, transistors, and other semiconductor devices. Here the ***qualitative properties of devices are determined by the shape of the electric field*** (Streetman, 1972; Sze, 1981).

This idea needs expansion because it may serve as a productive analogy to motivate future research on channels as practical devices. When the electric field has one shape, a transistor will behave like a resistor with linear $I(V)$ relations; for another shape of field, it will behave like a diode, with exponential $I(V)$ relations. For other, shapes of field, the transistor does more interesting things, acting as a linear amplifier, as a nonlinear logarithmic amplifier, as a multiplier, a limiter, and so on and so forth. A transistor can be many different devices without any physical change inside the transistor. The distribution of its fixed charge is not changed. All that is changed is the shape of the electric field and that is simply changed by applying different steady potentials (and currents) to the terminals of the devices. A great deal of our technology (and standard of living) arises because of the ease with which definite reproducible functions can be performed by transistors.

It is possible that proteins and channels exploit the *PNP* equations in the same way, although that is certainly only speculation at this time. What is clear is that assuming constant profiles of potential, and rate constants independent of concentration and shielding phenomena, makes it difficult to study that possibility.

**_What has been done_**? Channel permeation has historically been studied in two different traditions. Diffusion models of permeation started a long time ago, and reached many workers through the papers of Goldman, 1943, and Hodgkin and Katz, 1949, producing the widely used GHK equation. These papers use the diffusion equation to describe ion motion much as *PNP* does, but they allow the permeating ion to interact through the protein *only* through the diffusion coefficient, the permeability. The potential profile is assumed constant in space and with conditions, and *the protein is assumed to have no role in creating this profile*. Despite the evident historical importance of these papers, and the great advance they represented, they clearly are inadequate to describe how the charge and structure of a protein produces permeation. The relevant variables do not appear in the theory and so can have no effect on its outputs.

If the constant field theory is abandoned, we have the painful necessity of abandoning its consequences. Quantitative estimates of selectivity built on the permeability equation of constant field theory cannot be made using an equation without foundation. Otherwise, we abandon the most fundamental of principles of science, namely to derive theories from principles and physical models and then test those theories against experiments. If there is no sensible physical model that gives an equation, checking that equation against experiments has only limited use, and no physical meaning.

Unfortunately, a qualitative theory of permeability is not yet available to replace that of constant field theory. It is being worked on, but is not finished. Until then, we must work without the crutch of the GHK equation, in my opinion, since a broken crutch is likely to give way at the most awkward moment, casting its user into the maelstrom and turning a cripple into a dead man.

Permeation has also been studied in the tradition of barrier models, in the spirit of Eyring models of chemical reactions. This subject has been extensively reviewed and that need not be repeated here (Cooper, Jakobsson, and Wolynes, 1985; Cooper, Gates and Eisenberg, 1988a; Cooper, Gates and Eisenberg, 1988b; Nonner and Eisenberg, 1998). Suffice it to say that such models have certain difficulties:

1) The models assume large barriers. They do not apply if barriers are small.

2) The models assume barriers that do not change with concentration or applied potential.

3) The models do not involve a physical description of the channel protein. They do not specify how the potential barrier arises from the charge and/or chemical properties of the protein.

4) The models use a form of barrier theory applicable to gas phase chemical reactions even though ion permeation occurs in a condensed phase. That is to say, they describe the rate constant by $k = (k_B T/h)\exp(\Phi e/k_B T)$. If the correct expression (14) is used (Fleming, Courtney and Balk, 1986; Hänggi, Talkner and Borokovec, 1990; Pollak, 1996), *barrier models cannot predict currents of the same order of magnitude as observed in most channels*.

5) In many cases, barrier models have been used arbitrarily, with any prefactor the scientist wishes, even with prefactors that change from condition to condition. In this case, barrier models do not form a physical theory and so cannot be used to relate structure to function.

For these reasons, I believe that barrier models of the type widely used in the literature of permeation must be abandoned. What to replace them with is an open question. I believe *PNP* is a decent first replacement. But it is obviously only a crude approximation to the real physics of permeation and needs to be extended and replaced with high resolution models.

**_What needs to be done_**? Much must be done to create a new model of permeation. Clearly, the model must be selfconsistent, computing the electric field from all the charges present; clearly it must have high resolution and predict the actual parameters measured and controlled in experiments. *PNP* does much of this, but with evident difficulties.

*PNP* in the simple form presented here does not deal with selectivity between ions. *PNP* can be extended to do this by considering the excluded volume of ions, by describing the chemical potential of solutions of spheres. This work has just begun (Nonner, Catacuzzeno, and Eisenberg, 2000), but the beginning is hopeful (Boda*, et al.,* 2000). Many mean field theories exist of concentrated salt solutions; many experimental measurements of the excess chemical potential of these solutions have been made. It is already clear that simply including these excess chemical potentials is enough to give selectivity of the type found in L-type calcium channels. Much work is needed to see how far this approach can be extended.

*PNP* does not deal with unidirectional fluxes in its simplest form, although it can be extended to do so. But unidirectional fluxes are important in defining single file properties of channels. A selfconsistent theory of unidirectional fluxes is clearly needed.

Finally, *PNP* is a mean field theory, ready from its conception to be replaced by a trajectory based theory, in which individual atoms and trajectories are computed and analyzed. The difficulties of such a theory should not be minimized. For example, the experience of hundreds or thousands of physicists show how easy it is to make subtle errors in the calculations. The general rule is that a trajectory based calculation must be shown explicitly to give Ohm's law on the one hand and on the other hand, to calculate the electrostatic energy of a simple geometry (i.e., its electric field or capacitance) before it can be accepted as correctly formulated and programmed.

There is no doubt that such a theory will reveal inadequacies in *PNP*, because there is no doubt that a single mean field theory cannot describe the full domain of behavior of channels, e.g., when discreteness of charge matters. This has been evident to the authors of *PNP* since before its conception. Indeed, these difficulties nearly blocked the romance needed to make conception possible. What is surprising is not the inadequacies of *PNP* but rather its successes. As in the world of semiconductors, a mean field theory has helped considerably in understanding the transport of charge, the permeation of ions through channels.

As we move to higher resolution selfconsistent models, let us hope we can retain the virtues of simplicity and not be forced to swim in a sea of trajectories, in which we can drown unknowing of the buoyancy simple physics can provide us, when properly used.

## ACKNOWLEDGEMENT

The calculations shown in the Figures were performed by Uwe Hollerbach as part of our analysis of gramicidin (Hollerbach, *et al.,* 2000). It is a pleasure to thank Uwe, and our co-authors, David Busath and Duan Chen for their contributions to this paper. I am grateful to Boaz Nadler for advice concerning the Langevin equation. Lou DeFelice suggested I write this paper and made continual most helpful criticisms. Without him, it would not exist.